
\magnification=1200
\vsize=7.5in
\hsize=5.6in
\tolerance 500

\def\thru#1{\mathrel{\mathop{#1\!\!\!/}}}
\def\uprightarrow#1{\raise2ex\hbox{$\rightarrow$}\mkern-16.5mu #1}
\def\upleftarrow#1{\raise2ex\hbox{$\leftarrow$}\mkern-16.5mu #1}

\def\footnoterule{\kern-3pt \hrule width \hsize \kern2.6pt}
\pageno=0
\footline={\ifnum\pageno>0 \hss --\folio-- \hss \else\fi}

\baselineskip 15pt plus.5pt minus.5pt

\centerline{{\bf CHIRAL-ODD AND SPIN-DEPENDENT}\footnote{*}
{This work is
supported in part by funds provided by the U.S. Department of Energy (D.O.E.)
under contract \#DE-AC02-76ER03069.}}
\centerline{\bf QUARK FRAGMENTATION FUNCTIONS AND THEIR APPLICATIONS}
\vskip 24pt
\centerline{Xiangdong Ji}
\vskip 12pt
\centerline{\it Center for Theoretical Physics}
\centerline{\it Laboratory for Nuclear Science and Department of Physics}
\centerline{\it Massachusetts Institute of Technology}
\centerline{\it Cambridge, Massachusetts 02139}
\vfill
\centerline{\bf ABSTRACT}
\bigskip

\midinsert
\baselineskip 24pt plus 1pt minus 1pt
\narrower
\narrower
\noindent
We define a number of quark fragmentation functions for spin-0, -1/2 and -1
hadrons, and classify them according to their twist, spin and chirality.  As
an example of their applications, we use them to analyze semi-inclusive
deep-inelastic scattering on a transversely polarized nucleon.
\endinsert  

\vfill
\centerline{Submitted to: {\it Phys. Rev. D}}
\vfill
\line{CTP\#2219 \hfil hep-ph/9307235 \hfil June 1993}
\eject

\goodbreak\bigskip
\noindent{\bf I. INTRODUCTION}
\medskip

In high-energy processes, the
structure of hadrons is described by parton
distributions, or in broader sense, parton correlations.
In previous work$^{1-5}$, we have introduced
and exploited a number of low-twist
parton distributions, with some producing novel
spin-dependent and chiral-flip effects
in hard scattering processes.
These processes in turn allow us to gain
access to these distributions
experimentally and thereby help us to
learn the non-perturbative QCD physics of
hadrons. Among the distributions that
we have discussed, the quark
transversity distribution in the nucleon, which
is defined by the following light-cone correlation,$^{1,6}$
$$     h_1(x) = {1\over 2} \int {d \lambda \over 2\pi}
      e^{i\lambda x} \langle PS_\perp|
          \bar \psi(0)\thru n \gamma_5 \thru S_\perp
     \psi(\lambda n) |PS_\perp \rangle, \eqno(1) $$
is particularly interesting: it is one of the
three distributions which characterize
the state of quarks in the nucleon in
the leading-order high-energy processes;
its unweighted sum rule measures the {\it tensor
charge} of the nucleon, which is identical to the
axial charge in non-relativistic quark
models; it is a chiral-odd distribution (containing
both left and right handed quark fields)
so it does not appear in many well-known
inclusive hard processes such as deep-inelastic scattering.
Because of this last feature, we find it
cannot be easily measured experimentally.

To see what criteria an underlying
physical process has to meet in order to
measure the transversity distribution,
we consider the so-called ``cut diagrams''
for the {\it cross section} of the process,
which are obtained by gluing together the
Feynman diagrams for the amplitude and its complex
conjugate. In a cut diagram, a quark flowing out of a
hadron will come back to it after a series of scattering.
For $h_1(x)$ to appear, the chirality of the
quark must be flipped when it returns. This occurs
if the quark goes through some soft processes during scattering,
as shown in Fig. 1a. The only exception, a hard
process which flips chirality, is a mass insertion,
shown in Fig. 1b. For the light ($u$ or $d$) quarks,
the mass insertion is suppressed by $m/\Lambda_{\rm QCD}$
and is ignorable. [Mass insertions might be significant
for heavy quarks but they are not the subject of this paper.]
Chirality can be flipped in a parton distribution
as in the Drell-Yan process shown in Fig. 1c, where
the quark line goes through the interior of another hadron,
or in a quark fragmentation process in hadron production
shown in Fig. 1d, where the quark line goes through
a fragmentation vertex). To measure the transversity
distribution utilizing the second mechanism,
we must clarify the structure of
fragmentation vertices.

The semi-inclusive hadron production from a quark fragmentation
is described by fragmentation functions.
As is shown in ref. 7, parton fragmentation
functions in QCD are defined as matrix
elements of quark and gluon field operators
at light-cone separations. Thus, their
twist, spin, and chirality structures
shall be as rich as parton distribution
functions. In particular, there shall be
a correspondent fragmentation function for each
parton distribution function defined in ref. 2.
As we shall show blow, there also
exit additional fragmentation functions
due to hadron final state interactions.
Despite their similarity, fragmentation functions
are more difficult to calculate than
distribution functions.
However, our purpose here is to define them
and to study the circumstances under which
they contribute to scattering processes.

This paper is organized as follows. In section II,
we introduce complete twist-two and -three and
a part of twist-four quark fragmentation functions
for production of spin-0, -1/2, and -1 hadrons.
In section III, we study measurement of the nucleon's
transversity distribution in deep-inelastic scattering
using the chiral-odd quark fragmentation functions
defined in section II. We specifically
consider three hadron-production processes:
single-pion production, spin-1/2 baryon production,
and vector meson and two-pion production.
The first process uses polarized beam and target, and
the double spin asymmetry vanishes in high-energy
limit. The second process uses unpolarized lepton beam, however,
requires measurement of the
spin-polarization of the produced baryons. The third process
is a single-spin process, which utilizes a fragmentation
function arising from hadron final state
interactions. We conclude
the paper in section IV.

\goodbreak\bigskip
\noindent{\bf II. QUARK FRAGMENTATION FUNCTIONS}
\medskip

Quark fragmentation functions were
introduced by Feynman
to describe hadron production
from the underlying hard parton processes.$^8$
In QCD, it is possible to obtain
analytical formulas for these functions
in terms of the matrix elements of the quark and gluon fields
as for quark distributions
in a hadron.$^7$ In addition to
the well-known spin-independent, chiral-even fragmentation
function $D(z)$ (we shall call it $\hat f_1(z)$)
widely discussed in literature,
one can introduce various chiral-odd and
spin-dependent fragmentation functions,
which are capable of producing novel effects in lepton-hadron
and hadron-hadron scattering.$^9$ In this section,
we define fragmentation functions involving quark
bilinears for production of spin-0, -1/2, and -1 hadrons.
The discussion here can be easily generalized to
gluons and more complicated fragmentation processes.

\goodbreak\medskip
\noindent{\bf 1. Fragmentation Functions For Spin-0 Meson}
\smallskip

Let us consider pion production, or equivalently,
production of any hadron
whose spin is not observed. In this case, generalizing the
procedure in refs. 2 and 7, we can define three fragmentation
functions with quark fields alone,
$$       z\int {d\lambda \over 2\pi}
           e^{-i\lambda/ z}
       \langle 0|\gamma^{\mu}\psi(0)|\pi(P)X\rangle
       \langle \pi(P) X|\bar \psi(\lambda n)|0\rangle
          = 4[\hat f_1(z) p^{\mu} + \hat f_4(z) M^2n^{\mu}],
          \eqno(2) $$
$$       z\int {d\lambda \over 2\pi}
           e^{-i\lambda / z}
       \langle 0|\psi(0)|\pi(P)X\rangle
       \langle \pi(P) X|\bar \psi(\lambda n)|0\rangle
          = 4M \hat e_1(z),
         \eqno(3) $$
where $P$ is the four-momentum of the pion and $p$ and
$n$ are two light-like vectors such that
$p^2=n^2=0$, $p^-=n^+=0$, $p\cdot n =1$, and $P=p + nm_\pi^2/2$.
All Dirac indices on quark fields are implicitly contracted.
[Our notation for fragmentation functions is analogous to
the notation for distribution functions developed in refs. 2 and 5,
the caret denotes fragmentation.]
The mass $M$ is a generic QCD mass scale, and we
avoid use of the produced hadron mass because of
the singular behavior introduced in the chiral limit (the
left hand side of (3) does not vanish as $m_\pi\rightarrow 0$).
The summation over $X$ is implicit and covers
all possible states which can be populated by the
quark fragmentation. The state $|\pi(P)X\rangle$ is
an incoming scattering state between $\pi$ and $X$.
The renormalization point ($\mu^2$) dependence is suppressed
in (2) and (3). QCD radiative corrections induce log$\mu^2$
dependence in the fragmentation functions, which is compensated by the
log$Q^2/\mu^2$-dependence of their coefficients
in expressions for observed cross sections. The resulting
log$Q^2$ dependence, or the Alteralli-Parisi evolution,$^{10}$
is an important aspect of fragmentation
processes which we put aside while we classify
their spin and chirality properties. Here we work in $n\cdot A = 0$
gauge, otherwise gauge links have to be added to ensure
the color gauge invariance.
{}From a simple dimensional analysis,
we see that $\hat f_1(z)$, $\hat e_1(z)$, and $\hat f_4(z)$
are twist-two, -three, and -four, respectively; and
from their $\gamma$-matrix structure,
$\hat f_1(z)$ and $\hat f_4(z)$ are chiral-even
and $\hat e_1(z)$ is chiral-odd.
Hermiticity guarantees
these fragmentation functions are real.

The chiral-odd fragmentation
function $\hat e_1(z)$ involves both ``good'' and ``bad''
components of quark fields on the light cone
($\bar\psi\psi = \bar\psi_+\psi_- +
\bar\psi_-\psi_+$, where $\psi_{\pm} = P_{\pm}\psi$
with $P_{\pm} = {1\over 2}\gamma^{\mp}\gamma^{\pm})$.
Using the QCD equation of motion (neglecting
the masses for light quarks),
$$     i{d\over d\lambda} \psi_-(\lambda n)
         = -{1\over 2}\thru n i{\thru {\uprightarrow D}}_\perp(\lambda n)
        \psi_+(\lambda n),  \eqno(4)                           $$
where ${\uprightarrow D}_\perp^\alpha = {\uprightarrow D}^\alpha
      - {\uprightarrow D}\cdot np^\alpha + {\uprightarrow D}\cdot pn^\alpha $
and $i{\uprightarrow D}^\alpha(\lambda n)
= i{\uprightarrow \partial}^\alpha -gA^\alpha(\lambda n)$.
we rewrite $\hat e_1(z)$ in (3) as,
$$ \eqalign{      \hat e_1(z)  = - {z^2\over 8 M}
            \int {d\lambda \over 2\pi} e^{-i\lambda/z}
    \Big[ & \langle 0|\thru n i\thru {\uprightarrow D}_\perp(0) \psi_+(0)
              |\pi(P)X\rangle \langle \pi(P)X|
            \bar \psi_+(\lambda n)|0\rangle \cr
           +&  \langle 0| \psi_+(0)
              |\pi(P)X\rangle \langle \pi(P)X|
            \bar \psi_+(\lambda n)\thru n i\thru
{\upleftarrow D}_\perp(\lambda n) |0\rangle \Big].  } \eqno(5) $$
where $i{\upleftarrow D}^\alpha(\lambda n)
= i{\upleftarrow \partial}^\alpha +gA^\alpha(\lambda n)$.
Thus the twist-three
fragmentation explicitly involves three parton fields:
two quark and one gluon. The appearance of eq. (5)
motivates us to introduce a fragmentation density
matrix,
$$  \eqalign{ \hat M^\alpha_{\rho\sigma}(z, z_1)
       = &  \int {d\lambda\over 2\pi} {d\mu\over 2\pi}
             e^{-i\lambda /z} e^{-i\mu(1/z_1 - 1/z)}
   \langle 0|i{\uprightarrow D}^\alpha_\perp(\mu n)\psi_\rho(0)|\pi(P)X\rangle
             \langle\pi(P) X|\bar \psi_\sigma(\lambda n) |0\rangle \cr &
        +  \int {d\lambda\over 2\pi} {d\mu\over 2\pi}
             e^{i\lambda /z} e^{i\mu(1/z_1 - 1/z)}
          \langle 0|\psi_\rho(\lambda n)|\pi(P)X\rangle
             \langle\pi(P) X|\bar \psi_\sigma(0)i{\upleftarrow
         D}^\alpha_\perp(\mu n)  |0\rangle. }
        \eqno(6)  $$
where $\alpha$ is restricted to transverse dimensions.
It has the following expansion
in the Dirac spin space,
$$         \hat  M^\alpha(z, z_1) = M
               \gamma^\alpha \thru p {\hat E(z, z_1)\over z} + ... \eqno(7)  $$
where $\hat E(z, z_1)$ is a real, chiral-odd fragmentation
function
involving two light-cone fractions and the ellipsis denotes
higher-twist contributions. The function $\hat E(z,z_1)$ can
be isolated from $\hat M^\alpha$ through a projection:
$\hat E(z,z_1) = z/(8M)
{\rm Tr}\thru n \gamma_\alpha \hat M^\alpha(z,z_1)$.
{}From eqs. (5)-(7), it is easy to prove,
$$             \hat e_1(z) = - z \int \hat E(z, z_1)
           d\big({1\over z_1}\big). \eqno(8)  $$
Therefore, $\hat e_1(z)$ is just a special moment of
$\hat E(z, z_1)$. As a consequence, a measurement of
$\hat e_1(z)$ at one momentum scale is not
sufficient to determine its value at other scales,
because an Alteralli-Parisi type of evolution
equation exists only for $\hat E(z,z_1)$,
not for a subset of its moments.$^{11}$
This property of $\hat e_1(z)$
contrasts that of twist-two fragmentation functions,
such as $\hat f_1(z)$.

\goodbreak\medskip
\noindent
{\bf 2. Fragmentation Functions Arising From Hadron Final-State Interactions}
\smallskip

The quark fragmentations introduced above have
a one-to-one correspondence with the quark
distributions introduced for a spin-0 meson.
In practice, one can define one additional
fragmentation function for the pion,
$$   z\int {d\lambda \over 2\pi}
           e^{-i\lambda/ z}
       \langle 0|\sigma^{\mu\nu}i\gamma_5\psi(0)|\pi(P)X\rangle
       \langle \pi(P) X|\bar \psi(\lambda n)|0\rangle
        = 4M\epsilon^{\mu\nu\alpha\beta}p_\alpha n_\beta
         \hat e_{\bar 1}(z)    \eqno(9)
          $$
If there were no final state interactions between $\pi$ and
$X$, the state $|\pi(P)X\rangle$ transforms as a free state
under time-reversal symmetry and $\hat e_{\bar 1}(z)$
vanishes identically. Thus the magnitude of $\hat e_{\bar 1}(z)$
depends crucially on the effects of hadron final
state interactions.

To illustrate that such fragmentation functions do
exit, we consider production of an electron-positron
pair from a virtual photon of mass $4m^2_e <q^2<16m^2_e$
in Quantum Electrodynamics. The production cross section
is proportional to the vacuum tensor,
$$     W^{\mu\nu} = {1\over 2\pi} \int d^4x e^{iqx}
            \langle 0|J^{\mu}(x)|e^+(P)e^-\rangle
            \langle e^+(P)e^-|J^{\nu}(0)|0\rangle \eqno(10)   $$
And this, according to Lorentz invariance, has
the following decomposition in terms of Lorentz
scalers,
$$     W^{\mu\nu} = (-g^{\mu\nu} + {q^\mu q^\nu\over q^2})
W_1 + ... + i(P^\mu q^\nu - P^\nu q^\mu)W_6  \eqno(11) $$
If neglecting the final state interactions
between the electron and positron, one can prove
immediately $W_6=0$ due to time-reversal invariance.

However, if taking into account one-photon exchange,
one finds,
$$     W_6 = C\sqrt{1-4m^2_e/q^2} \theta(q^2-4m^2_e) \eqno(12) $$
where $C$ is an unimportant numerical constant. The $\theta$ function
indicates the final state interaction vanishes if $q^2<4m^2_e$,
in particular, if $q^2<0$, $W^{\mu\nu}$ is proportional
to the photon-electron scattering cross section,
to which we know $W_6$ does not contribute.

The fragmentation function $\hat e_{\bar 1}(z)$ is chiral-odd
and twist-three. It is intimately related to $\hat e_1(z)$
introduced in the previous subsection. It is simple to
show that it contributes to $W_6$ type of terms
in semi-inclusive production of hadrons
in $e^+e^-$ annihilation.

\goodbreak\medskip
\noindent
{\bf 3. Fragmentation Functions For Spin-1/2 Baryon}
\smallskip

Now we turn to consider the quark fragmentation for
a spin-1/2 baryon. Eight
more fragmentation functions
can be introduced through bilinear quark fields besides
these in eqs. (2), (3) and (9).
They all depend on the polarization of the baryon:
four of them are related to the
longitudinal polarization and the other four to the
transverse polarization,
$$ \eqalign{  &    z\int {d\lambda \over 2\pi}
           e^{-i\lambda /z}
       \langle 0|\gamma^{\mu}\gamma_5\psi(0)|B(PS)X\rangle
       \langle B(PS) X|\bar \psi(\lambda n)|0\rangle   \cr &
          = 4\Big[\hat g_1(z) p^{\mu}(S_{||}\cdot n)
          + M\hat g_T(z) S^\mu_\perp +
           M^2\hat g_3(z)(S_{||}\cdot n) n^\mu \Big] } \eqno(13)
          $$
$$ \eqalign{    &  z\int {d\lambda \over 2\pi}
           e^{-i\lambda/ z}
       \langle 0|\sigma^{\mu\nu}i\gamma_5\psi(0)|B(PS)X\rangle
       \langle B(PS) X|\bar \psi(\lambda n)|0\rangle
         \cr &  = 4\Big[\hat h_1(z)(S_{\perp}^{\mu}p^{\nu} -
           S_{\perp}^{\nu}p^{\mu} )
          + \hat h_L(z) M (p^\mu n^\nu - p^\nu n^\mu)(S_{||}\cdot n)
         \cr & +  \hat h_3(z)M^2 (S_{\perp}^{\mu}n^{\nu} -
           S_{\perp}^{\nu}n^{\mu})  + ...\Big], }   \eqno(14)
          $$
$$ \eqalign{  &    z\int {d\lambda \over 2\pi}
           e^{-i\lambda /z}
       \langle 0|\gamma^{\mu}\psi(0)|B(PS)X\rangle
       \langle B(PS) X|\bar \psi(\lambda n)|0\rangle   \cr &
          = 4\Big[\hat g_{\bar T}(z)MT^\mu_\perp + ...\Big]} \eqno(15)
          $$
$$ \eqalign{  &    z\int {d\lambda \over 2\pi}
           e^{-i\lambda /z}
       \langle 0|\gamma_5\psi(0)|B(PS)X\rangle
       \langle B(PS) X|\bar \psi(\lambda n)|0\rangle   \cr &
          = 4 M\hat h_{\bar L}(z) (S_{||}\cdot n)  }  \eqno(16)
          $$
where dots denote terms already appeared in (2), (3) and (9),
$B(PS)$ represents the spin-1/2 baryon with the
four-momentum $P$ and polarization $S$ (we
write $S^\mu = S\cdot n p^\mu
+ S\cdot pn^\mu  + M_BS^\mu_\perp$ with the baryon mass $M_B$),
and $T^\mu = \epsilon^{\mu\nu\alpha\beta}S_{\perp\nu}p_\alpha
n_\beta$ is a transverse vector orthogonal to $S_\perp^\mu$.
Again, through dimensional analysis,
$\hat g_1(z)$ and $\hat h_1(z)$ are twist-two;
$\hat g_T(z)$, $\hat h_L(z)$, $\hat g_{\bar T}(z)$ and
$\hat h_{\bar L}(z)$ are twist-three;
and $\hat g_3(z)$ and $\hat h_3(z)$ are twist-four.
The fragmentation functions $\hat g_{\bar T}(z)$ and
$\hat h_{\bar L}(z)$ vanish identically if without
final-state interactions.

As was the case for $\hat e_1(z)$, at the level of twist-three,
$\hat g_T(z)$ and $\hat h_L(z)$ are not the most
general fragmentation functions.
Using (4), we derive,
$$\eqalign{         \hat g_T(z)  =&  - {z^2\over 8M}
            \int {d\lambda \over 2\pi} e^{-i\lambda/z}
 \langle 0|i{\uprightarrow D}_\perp(0)\cdot S_\perp\thru n\gamma_5 \psi_+(0)
              |B(PS_\perp)X\rangle \cr &
            \times \langle B(PS_\perp)X|
             \bar \psi_+(\lambda n)|0\rangle  \cr &
               + {z^2\over 8M}
            \int {d\lambda \over 2\pi} e^{-i\lambda/z}
            \langle 0|{\uprightarrow D}_\perp(0)\cdot T_\perp\thru n \psi_+(0)
              |B(PS_\perp)X\rangle \cr &
            \times \langle B(PS_\perp)X|
             \bar \psi_+(\lambda n)|0\rangle
             + {\rm C.C.}  }       \eqno(17)   $$
$$\eqalign {     \hat h_L(z)  = &  - {z^2\over 8M}
            \int {d\lambda \over 2\pi} e^{-i\lambda/z}
            \langle 0|
             i \thru {\uprightarrow D}_\perp(0)\thru n \gamma_5 \psi_+(0)
              |B(PS_{||})X\rangle     \cr &
             \times \langle B(PS_{||})X|
             \bar \psi_+(\lambda n)|0\rangle
             + {\rm C.C.}}   \eqno(18)    $$
where ${\rm C.C.}$ stands for complex conjugate.
The generalization of eqs. (17) and (18) to
two-light-cone-fraction distributions can be made by
defining $\hat M^\alpha (z,z_1)$ for
the baryon just as for the pion in eq. (6). In addition, we
need to define a new fragmentation density matrix,
$$  \eqalign{ \hat N^\alpha_{\rho\sigma}(z, z_1)
       = &  - \int {d\lambda\over 2\pi} {d\mu\over 2\pi}
             e^{-i\lambda /z} e^{-i\mu(1/z_1 - 1/z)}
           \langle 0|i{\uprightarrow D}^\alpha_\perp(\mu n)
              \psi_\rho(0)|B(PS)X\rangle   \cr & ~~~\times
             \langle B(PS) X|\bar \psi_\sigma(\lambda n) |0\rangle \cr &
        +  \int {d\lambda\over 2\pi} {d\mu\over 2\pi}
             e^{i\lambda /z} e^{i\mu(1/z_1 - 1/z)}
          \langle 0|\psi_\rho(\lambda n)|B(PS)X\rangle
             \langle B(PS) X|\bar \psi_\sigma(0)i{\upleftarrow
         D}^\alpha_\perp(\mu n)  |0\rangle. }
        \eqno(19)  $$
which is the same as $\hat M^\alpha$ except the minus sign for the
first term. Making expansion in the spin space, we have,
$$ \eqalign{        \hat  M^\alpha(z, z_1)  & = M
               \gamma^\alpha \thru p \hat E(z, z_1)/z + iMT^\alpha_\perp
              \thru p \hat G_1(z, z_1)/z + ...  \cr
             \hat  N^\alpha(z, z_1) & = MS^\alpha_\perp \gamma_5 \thru p
                  \hat G_2(z, z_1)/z
               + M\gamma^\alpha\thru p \gamma_5 \hat H(z, z_1)/z + ... }
            \eqno(20)  $$
The fragmentation functions can be projected
from the density matrices: $G_1 = iz/(4M)
{\rm Tr} \thru n\gamma_5 T_{\perp\alpha} M^\alpha$,
$G_2 = -z/(4M){\rm Tr} \thru n\gamma_5 S_{\perp\alpha} N^\alpha$,
and $H = -z/(8M){\rm Tr} \gamma_{\perp\alpha}
\thru n\gamma_5 N^\alpha$. It is easy to prove that
$$            \hat g_T(z) = - {z\over 2} \int
                  d({1\over z_1})\Big [  \hat G_1(z, z_1)
                   + \hat G_2(z, z_1)\Big],     \eqno(21)        $$
$$            \hat h_L(z) = - z \int
                  d({1\over z_1})\hat  H(z, z_1).
             \eqno(22) $$
These relations are useful for proving electromagnetic gauge invariance
of scattering amplitudes, as an example shows in section III.

\goodbreak\medskip
\noindent
{\bf 4. Fragmentation Functions For Spin-1 Meson}
\smallskip

Finally, we consider quark fragmentation
functions for vector meson production.
To facilitate counting, let us define the
quark-meson forward scattering amplitudes, $A_{hH;h'H'}$,
where $h(h')$ and $H(H')$ are quark and meson helicities,
respectively. The combination $A_{{1\over 2}1;{1\over 2}1}
+ A_{{1\over 2}-1;{1\over 2}-1}+A_{{1\over 2}0;{1\over 2}0}$
is independent of the meson polarization, from which we
define four fragmentation functions $\hat f_1$,
$\hat e_1$ and $\hat e_{\bar 1}$, and $\hat f_4$,
depending on what components of quark
fields form the amplitude: good-good ($++$), good-bad ($+-$),
or bad-bad ($--$). Of course, they are
what we have just defined in (2), (3) and (9).
Similarly, the combination $A_{{1\over 2}1;{1\over 2}1}
+ A_{{1\over 2}-1;{1\over 2}-1}-2A_{{1\over 2}0;{1\over 2}0}$
depends on the LL type of tensor polarization of the meson
(see below for definition) and the corresponding four
fragmentation functions are $\hat b_1$, $\hat b_2$
and $\hat b_{\bar 2}$, and $\hat b_3$;
the combination $A_{{1\over 2}1;{1\over 2}1}
- A_{{1\over 2}-1;{1\over 2}-1}$ depends on
the TT type of vector polarization and the
associated fragmentation functions are $\hat g_1$,
$\hat h_2$ and $\hat h_{\bar 2}$, and $\hat g_3$;
the combination $A_{{1\over 2}0;-{1\over 2}1}
- A_{-{1\over 2}1;{1\over 2}0}$ is related to
the LT type of vector polarization
and the associated fragmentation functions are defined
as $\hat h_1$, $\hat g_2$ and $\Delta \hat g_{\bar 2}$, and
$\hat h_3$; and finally, the combination
$A_{{1\over 2}0;-{1\over 2}1}
+ A_{-{1\over 2}1;{1\over 2}0}$ is related to
the LT type of tensor polarization and the associate
fragmentation functions are defined as
$\hat h_{\bar 1}$, $\hat g_{\bar 2}$,
$\Delta \hat g_2$, and $\hat h_{\bar 3}$.
The spin and twist structure of these twenty fragmentation
functions are shown in Table 1, and the ones with
bar on their subscripts arising from
hadron final state interactions.

\midinsert
\bigskip
\baselineskip 15pt
\def\strut{\hbox{\vrule height 4pt depth 4pt width 0pt}}
\centerline{{\bf Table I}.~~~Quark fragmentation functions for vector mesons.}
\centerline
{Note: the functions with bar vanish if there are no final state interactions.}
\vskip-12pt
$$\vbox{\offinterlineskip
\hrule
\halign{
&\vrule#&\enskip\hfil$#$\hfil\enskip
&\vrule#&\enskip\hfil$#$\hfil\enskip
&\vrule#&\enskip\hfil$#$\hfil\enskip
&\strut#&\enskip\hfil$#$\hfil\enskip
&\vrule#&\enskip\hfil$#$\hfil\enskip
&\vrule#&\enskip\hfil$#$\hfil\enskip\cr
height5pt&&&\omit&&\omit&&\omit&&\omit&&\omit&\cr
&&&${\rm Twist-2}$&&\multispan3\hfil{\rm Twist-3}\hfil
&&${\rm Twist-4}$&&&\cr
&&&&&&&&&&&${\rm meson}$&\cr
&&&++&&+-${\rm (S)}$&&+-${\rm (A)}$&&--&&${\rm polarization}$&\cr
height5pt&&&&&&&&&&&&\cr
\noalign{\hrule}
height4pt&&&\omit&&\omit&&\omit&&\omit&&\omit&\cr
& A_{{1\over2}{1} \to {1\over2}{1}}
+ A_{{1\over2}{-1} \to {1\over2}{-1}}
+ A_{{1\over2}{0} \to {1\over2}{0}}
&& \skew4\hat f_1
&& \hat e_1
&& \hat e_{\bar{1}}
&& \skew4\hat f_4
&& {\rm S} &\cr
height4pt&\omit&&\omit&&\omit&&\omit&&\omit&&\omit&\cr
\noalign{\hrule}
height4pt&\omit&&\omit&&\omit&&\omit&&\omit&&\omit&\cr
& A_{{1\over2}{1} \to {1\over2}{1}}
+ A_{{1\over2}{-1} \to {1\over2}{-1}}
- 2A_{{1\over2}{0} \to {1\over2}{0}}
&& \hat b_1
&& \hat b_1
&& \hat b_{\bar{2}}
&& \hat b_3
&& {\rm T}_{\!LL} &\cr
height4pt&\omit&&\omit&&\omit&&\omit&&\omit&&\omit&\cr
\noalign{\hrule}
height4pt&\omit&&\omit&&\omit&&\omit&&\omit&&\omit&\cr
& A_{{1\over2}{1} \to {1\over2}{1}}
- A_{{1\over2}{-1} \to {1\over2}{-1}}
&& \hat g_1
&& \hat h_2
&& \hat h_{\bar{2}}
&& \hat g_3
&& {\rm V}_{\!TT} &\cr
height4pt&\omit&&\omit&&\omit&&\omit&&\omit&&\omit&\cr
\noalign{\hrule}
height4pt&\omit&&\omit&&\omit&&\omit&&\omit&&\omit&\cr
& A_{{1\over2}{0} \to -{1\over2}{1}}
- A_{-{1\over2}{1} \to {1\over2}{0}}
&& \hat h_1
&& \hat g_2
&& \Delta \hat g_{\bar{2}}
&& \hat h_3
&& {\rm V}_{\!LT} &\cr
height4pt&\omit&&\omit&&\omit&&\omit&&\omit&&\omit&\cr
\noalign{\hrule}
height4pt&\omit&&\omit&&\omit&&\omit&&\omit&&\omit&\cr
& A_{{1\over2}{0} \to -{1\over2}{1}}
+ A_{-{1\over2}{1} \to {1\over2}{0}}
&& \hat h_{\bar{1}}
&& \hat g_{\bar{2}}
&& \Delta \hat g_{2}
&& \hat h_{\bar{3}}
&& {\rm T}_{\!LT} &\cr
height4pt&\omit&&\omit&&\omit&&\omit&&\omit&&\omit&\cr
\noalign{\hrule}}}$$
\vskip-12pt
\endinsert

Now we relate these fragmentation
functions to the matrix elements
of the bilinear quark operators.
Since the meson polarization vector
$\epsilon^\mu$ appears in bilinear form
in all the matrix elements,
we introduce a rank-two tensor
$T^{\mu\nu} = \epsilon^\mu\epsilon^{*\nu}$.
Its trace $T^\mu_{~\mu}=S$, antisymmetric
part $T^{[\mu\nu]} = \epsilon^\mu\epsilon^{*\nu}
- \epsilon^\nu\epsilon^{*\mu}$, and
traceless-symmetric part $T^{\{\mu\nu\}} = \epsilon^\mu \epsilon^{*\nu} +
\epsilon^\nu \epsilon^{*\mu} - (\epsilon\cdot\epsilon^*)
g^{\mu\nu}/2$ represent the scalar, vector, and tensor
polarization of the meson. Together
with $p_\mu$ and $n_\mu$, they can be
used to build various Lorentz structures
to expand the quark matrix elements.
The coefficients of the expansion, depending on the
polarization and dimension of the associated
structures, can be uniquely identified with the
fragmentation functions in Table 1.

To illustrate this, take the scalar polarization $S$, from
which one can form one scalar $S$, two vectors $Sp^\mu$ and $Sn^\mu$,
and one tensor $\epsilon^{\mu\nu\alpha\beta} p_\alpha n_\beta S$, and
the coefficients of these structure shall be
$\hat e_1$, $\hat f_1$ and $\hat f_4$,
and $\hat e_{\bar 1}$, respectively. For the case of
tensor polarization, consider the projection of
$T^{\{\mu\nu\}}$ in longitudinal directions,
$T_{\{\alpha\beta\}}p^\alpha n^\beta(=T)(n^\mu p^\nu + n^\nu p^\mu)$,
which characterizes the LL type of tensor polarization.
With this one can construct one scalar $T$, two vectors
$Tp^\mu$ and $Tn^\nu$, and one tensor $\epsilon^{\mu\nu\alpha\beta}
p_\alpha n_\beta T$ and their coefficients are $\hat b_2$,
$\hat b_1$ and $\hat b_3$, and $\hat b_{\bar 2}$, respectively.
Proceeding in this way, define
$A = i\epsilon^{\alpha\beta\gamma\delta} \epsilon_\alpha
\epsilon_\beta^*p_\gamma n_\delta $ to characterize the TT type of
vector polarization, $S^\mu_\perp = i\epsilon^{\mu\alpha\beta\gamma}
p_\alpha n_\beta T_{[\gamma\delta]}n^\delta$ the LT type of
vector polarization, and $T^\mu_\perp = \epsilon^{\mu\alpha\beta\gamma}
p_\alpha n_\beta T_{\{\gamma\delta\}}n^\delta$ the
LT type of tensor polarization, and construct all possible
structures with them. The complete
expansion of quark matrix elements reads,
$$ \eqalign{      & z\int {d\lambda \over 2\pi}
           e^{-i\lambda/ z}
       \langle 0|\gamma^{\mu}\psi(0)|V(P\epsilon)X\rangle
       \langle V(P\epsilon) X|\bar \psi(\lambda n)|0\rangle  \cr
        &   = 4\Big[\hat f_1(z)Sp^{\mu} +\hat f_4(z)M^2Sn^{\mu} +
              \Delta \hat g_{\bar 2}(z)MiT^{[\mu\nu]}_\perp n_\nu  \cr &
          + \Delta \hat  g_2(z)M T^{\{\mu\nu\}}_\perp n_\nu
          +\hat b_1(z)Tp^{\mu} +\hat b_4(z)M^2Tn^{\mu} \Big], }
          \eqno(23) $$
$$       z\int {d\lambda \over 2\pi}
           e^{-i\lambda / z}
       \langle 0|\psi(0)|V(P\epsilon)X\rangle
       \langle V(P\epsilon) X|\bar \psi(\lambda n)|0\rangle
          = 4M\Big[S \hat e_1(z) + T\hat b_2(z)\Big],
         \eqno(24) $$
$$      z \int {d\lambda \over 2\pi}
           e^{-i\lambda / z}
       \langle 0|i\gamma_5\psi(0)|V(P\epsilon)X\rangle
       \langle V(P\epsilon) X|\bar \psi(\lambda n)|0\rangle
          = 4MA h_{\bar 2}(z),
         \eqno(25) $$
$$ \eqalign{  &    z\int {d\lambda \over 2\pi}
           e^{-i\lambda /z}
       \langle 0|\gamma^{\mu}\gamma_5\psi(0)|V(P\epsilon)X\rangle
       \langle V(P\epsilon) X|\bar \psi(\lambda n)|0\rangle   \cr &
          = 4\Big[\hat g_1(z)Ap^{\mu}
          + M\hat g_2(z) S^\mu_\perp + M\hat g_{\bar 2}(z) T^\mu_\perp +
           M^2\hat g_3(z)An^\mu \Big], } \eqno(26)
          $$
$$ \eqalign{    &  z\int {d\lambda \over 2\pi}
           e^{-i\lambda/ z}
       \langle 0|\sigma^{\mu\nu}i\gamma_5\psi(0)|V(P\epsilon)X\rangle
       \langle V(P\epsilon) X|\bar \psi(\lambda n)|0\rangle \cr &
          = 4\Big[\hat h_1(z)(S^\mu_\perp p^\nu - S^\nu_\perp p^\mu)
           + \hat h_{\bar 1}(z) (T^\mu_\perp p^\nu - T^\nu_\perp p^\mu)
          + \hat h_2(z)MA(p^\mu n^\nu - p^\nu n^\mu)\cr &
         +  \hat h_3(z)M^2 (S^\mu_\perp n^\nu - S^\nu_\perp n^\mu)
         + \hat h_{\bar 3}(z)M^2 (T^\mu_\perp n^\nu - T^\nu_\perp n^\mu) \cr &
         + e_{\bar 1}(z)M \epsilon^{\mu\nu\alpha\beta} p_\alpha n_\beta S
         + b_{\bar 2}(z)M \epsilon^{\mu\nu\alpha\beta} p_\alpha n_\beta T
              \Big]. }   \eqno(27)          $$
Thus, all the twenty fragmentation functions in Table 1
have expressions in QCD.

\goodbreak\bigskip
\hangindent=.5in \hangafter=1
\noindent
{\bf III. MEASURING THE TRANSVERSITY DISTRIBUTION
\hfil\break FROM DEEP-INELASTIC SCATTERING}
\medskip
\nobreak

As an example of applying the
fragmentation functions defined in the last section,
we consider measuring the nucleon's transversity
distribution $h_1(x)$ through deep-inelastic
scattering. Because $h_1(x)$
is chiral-odd, it does not appear in {\it inclusive}
deep-inelastic cross section if
the current quark masses are neglected.
However, $h_1(x)$ does appear in
semi-inclusive hadron production if
one takes into account the effects of the chiral-odd
fragmentation of the struck quark.
For pseudo-scalar meson
production, the leading chiral-odd quark
fragmentation function is
$\hat e_1(z)$, for spin-1/2 baryon production,
it is $\hat h_1(z)$, and for vector meson production,
it is $\hat h_{\bar 1}(z)$ or its generalization to
a fragmentation function for two pions.
In the following, we discuss these cases separately.

\goodbreak\medskip
\noindent
{\bf 1. Single-Pion Production}
\smallskip
\nobreak

We consider deep-inelastic scattering with
longitudinal polarized lepton on transversely
polarized nucleon target, focusing on pion production
in the current fragmentation region. Since
there is no {\it chiral-odd} twist-two fragmentation
function for the pion to couple with the $h_1(x)$
distribution in the nucleon, nor is there
{\it chiral-even} twist-two transverse-spin-dependent
distribution in the nucleon to
couple with the fragmentation function $\hat f_1(z)$
for the pion, the spin-dependent cross section
vanishes at the leading order in $Q$.
At the twist-three level (the order of $1/Q$),
$h_1(x)$ contributes through coupling
with the chiral-odd fragmentation function
$\hat e_1(z)$, and so does the chiral-even
transverse-spin distribution $g_T(x)$ through
the fragmentation function $\hat f_1(z)$.
Both contributions exist in Fig. 2a.
At the same order, we have to consider also
Figs. 2b and 2c, in which one radiative gluon
takes part in quark fragmentation,
and Figs. 2d and 2e, in which one gluon
from the nucleon participates in hard scattering.
These processes, representing coherent parton scattering,
introduce dependences on the
two-light-cone-fraction parton distributions,
$G_{1,2}(x,x_1)$, which are the parents of $g_T(x)$,$^2$
and fragmentation function, $\hat E(z,z_1$), which is the
parent of $\hat e_1(z)$.
However, as we shall show below,
they can be eliminated by using QCD equations of motion,
and the final result contains only $\hat e_1(z)$ and $g_T(x)$.

Let us first consider the contribution from the diagram
in Fig. 2a. Using the definition of the nucleon tensor,
$$  W_{\mu\nu} ={1\over 4\pi} \int e^{iq\cdot \xi}d^4\xi
          \langle PS|J_\mu(\xi)J_\nu(0)|PS\rangle,
       \eqno(28) $$
we obtain from this diagram,
$$     W^a_{\mu\nu} =
     {1\over 4\pi} \int {d^4k\over (2\pi)^4}
       \int {d^4P_\pi\over (2\pi)^4} 2\pi \delta(P_\pi^2-m_\pi^2)
       ~{\rm Tr} \Big[M_N(P,S_\perp,k) \gamma_{\mu}\hat M_\pi(k+q,P_\pi)
         \gamma_{\nu}\Big]  \eqno(29)  $$
where $m_\pi$ is the pion mass,
$$      M_N(P,S_\perp,k)_{\alpha\beta} =
               \int d^4\xi e^{i\xi\cdot k}
              \langle PS_\perp|\bar \psi_\beta(0)
            \psi_\alpha(\xi)|PS_\perp\rangle  \eqno(30)$$
is quark's spin-density matrix for the nucleon and
$$      \hat M_\pi(k,P_\pi)_{\alpha\beta} =
               \sum_X \int d^4\xi e^{-i\xi\cdot k}
              \langle 0|\psi_\alpha(0)|\pi(P_\pi)X\rangle
              \langle \pi(P_\pi)X|\bar \psi_\beta(\xi)|0\rangle
         \eqno(31)  $$
is the quark fragmentation density matrix for the pion.
Here $q$ is the four-momentum of the virtual photon,
and $P$ and $S_\perp$ are the nucleon's four-momentum and
polarization vectors, respectively.
We choose our coordinate system such
that $P = p + n M_N^2/2$,
$S_\perp = (0,1,0,0)$, and $q = -x_Bp + \nu n$,
where $p$ and $n$ are two light-like vectors defined in the last section,
$M_N$ is the mass of the nucleon, and $x_B$ is
the Bjorken scaling variable $x_B = Q^2/(2\nu)$.

To perform the $k$ integration in eq. (29), we make a
collinear expansion of quark momentum $k$ {\it along}
$p$ in the fragmentation density matrix,
$$        \hat M_\pi(k+q,P_\pi) = \hat M_\pi(k\cdot np+q, P_\pi)
               + (k-k\cdot n p)^\alpha {\partial
                 \hat M_\pi(k\cdot np+q, P_\pi) \over \partial k^\alpha}
         + ...   \eqno(32)  $$
We temporarily ignore the derivative term, whose
contribution will be combined with those from
Figs. 2d and 2e to form a gauge-invariant result.
The contribution to the $k$ integration from the
leading term in eq.(32) is,
$$         W^{a}_{\mu\nu}
              = {1\over 4\pi} \int {d^4P_\pi \over (2\pi)^4}
                 2\pi \delta(P_\pi^2 - m_\pi^2)
                \int dx {\rm Tr}\left[M_N(xp,S_\perp)\gamma_\mu
             \hat  M_\pi(xp+q, P_\pi) \gamma_\nu\right],  \eqno(33) $$
where the simplified quark spin-density matrix is
$$         M_N(xp,S_\perp)_{\alpha\beta}
           = \int { d\lambda \over 2\pi}
              e^{i\lambda x} \langle PS_\perp|
        \bar \psi_\beta(0) \psi_\alpha(\lambda n)
            |PS_\perp\rangle,   \eqno(34) $$
the structure of
which has been studied thoroughly in ref. 1.

To integrate out the transverse components of $P_\pi$
in eq. (33), we make a coordinate
transformation to a new system in
which $P_\pi$ and $q$ have only longitudinal components.
If we label momenta in the new system with prime, then,
to the order of our interest,
$$ \eqalign{ & P'^-_\pi = P^-_\pi, ~~P'^+_\pi = 0,
          ~~P'^i_\pi = 0 \cr &
  q'^- = P^-_\pi/z, ~~q'^+ = q^+, ~~q'^i = 0 \cr &
    xp'^-= 0, ~~xp'^+ = xp^+, xp'^i = -P_\pi^i/z} \eqno(35) $$
In the new system, $p'$ has non-vanishing
transverse components and as a consequence, the spin and
fragmentation density matrices in eq. (33) are now
linked through transverse-momentum integrations.
To decouple them, we Taylor-expand the spin-density
matrix,
$$     M_N(xp',S_\perp) = M_N(xp,S_\perp) -
          {P^i_\pi\over z} {\partial M_N(xp, S_\perp)\over \partial xp^i}
       + ...
        \eqno(36) $$
Here we have ignored the transverse components of $n'$, whose effects
are beyond twist-three. The contribution from
the derivative term in eq. (36) will be combined with those from
Figs. 2b and 2c to form a color gauge-invariance expression,
as is shown in eq. (43).
And the leading term contribution is,
$$         W^{a}_{\mu\nu}
              = {1\over 4\pi} \int dz
                \int dx 2\pi \delta ((xp+q)^2){\rm Tr}\left[M_N(xp,S_\perp)
             \gamma_\mu\hat  M_\pi(z,p_\pi/z) \gamma_\nu\right]  \eqno(37) $$
where the simplified fragmentation density matrix is
$$      \hat M_\pi(z,p_\pi/z)_{\alpha\beta} =
               \sum_X \int {d\lambda \over 2\pi}
                e^{-i\lambda /z}
              \langle 0|\psi_\alpha(0)|\pi(p_\pi)X\rangle
              \langle \pi(p_\pi)X|\bar \psi_\beta(\lambda n_\pi)|0\rangle
    \eqno(38) $$
Here we have neglected the pion mass and used two additional light-like
vectors $p_\pi$ and $n_\pi$ with $p_\pi = z\nu n$
and $p_\pi \cdot n_\pi = 1$.

Since spin asymmetry is our main interest, we
take the transverse-spin-dependent part of
the spin-density matrix from ref. 5,
$$  M_N(x,p,S_\perp) ={1\over 2} h_1(x) \gamma_5 \thru S_\perp\thru p
                   + {1\over 2} g_T(x) M\gamma_5 \thru S_\perp +... \eqno(39)$$
{}From eqs. (2) and (3), we have the
fragmentation density,
$$          \hat M_\pi(z)   = M {\hat e_1(z) \over z}
        + \thru p_\pi {\hat f_1(z)\over z}  + ...
        \eqno(40) $$
Substituting eqs. (39) and (40) into eq. (37) and simplifying the latter,
we have,
$$ \eqalign{    W^a_{\mu\nu} = {M\over 2\nu} &
             \Big[ \sum_a e_a^2h_1^a(x_B)\int \hat
             {e^a_1(z)\over z}dz ~~i\epsilon^{\mu\nu\alpha\beta}
              p_{\alpha}S_{\perp\beta}  \cr & +
              \sum_a e_a^2g_T^a(x_B)\int {\hat f^a_1(z)\over z}dz
          ~~i\epsilon^{\mu\nu\alpha\beta}
              S_{\perp\alpha}p_{\pi\beta} \Big].}  \eqno(41) $$
where the summation runs over different quark flavors
and their charge conjugation, and $e_a$ is electric charge
of quarks. As it stands, eq. (41) does not satisfy
electromagnetic gauge invariance, i.e., $W_{\mu\nu}q^\mu \ne 0$.

We turn to consider the contribution from Fig. 2b,
which involves an additional transversely-polarized gluon.
After the collinear expansion and coordinate transformation
discussed above, we find,
$$\eqalign {     W^b_{\mu\nu} = &
     {1\over 4\pi} \int dxdzd({1\over z_1})
            2\pi\delta((q+xp)^2)  \cr &
         ~\times {\rm Tr} \Big[M_N(xp,S_\perp) i\gamma_{\alpha}
          { i(x\thru p - (1/z-1/z_1)\thru p_\pi) \over
          (xp - (1/z-1/z_1)p_\pi)^2} \gamma_\mu
           \hat M^\alpha_1(z, z_1)
          \gamma_{\nu}\Big] } \eqno(42)    $$
where the fragmentation density matrix is,
$$  \eqalign{ \hat M^\alpha_1(z, z_1)_{\rho\sigma}
       = &  \int {d\lambda\over 2\pi} {d\mu\over 2\pi}
             e^{-i\lambda /z} e^{-i\mu(1/z_1 - 1/z)}
 \cr & \times \langle 0|i{\uprightarrow D}^\alpha_\perp(\mu n)\psi_\rho
               (0)|\pi(P_\pi)X\rangle
             \langle\pi(P_\pi) X|\bar \psi_\sigma
           (\lambda n) |0\rangle  }
        \eqno(43)  $$
The partial derivative in ${\uprightarrow D}^\alpha_\perp$ comes from the
collinear expansion for Fig. 2a as
explained after eq. (36).
Because the state $|\pi(P_\pi)X\rangle$
is an incoming scattering state which changes to
an outgoing scattering state after time reversal, $\hat M^\alpha_1(z, z_1)$
do not have a simple hermitian conjugation property.
As a consequence, if we make an expansion,
$\hat M_1^\alpha = M\gamma^\alpha \thru p \hat E_1(z,z_1) + ...$,
$E_1(z,z_1)$ is not a real quantity. However,
its imaginary part, which we are going to ignore,
contributes only to single-spin asymmetry.
Its real part is just $E(z,z_1)/2$, which was defined in the last
section. Inserting the expansion into eq. (42) and
using eq. (8) to eliminate $\hat E(z,z_1)$, we find,
$$     W^b_{\mu\nu} = {M\over 2\nu}
             \sum_a e_a^2 h_1^a(x)\int dz \hat e^a(z)
           {1\over x_B z^2}{1\over p\cdot p_\pi}
            p_\pi^{\mu}i\epsilon^{\nu\alpha\beta\gamma}
              S_{\perp\alpha} p_{\beta} p_{\pi\gamma}.  \eqno(44) $$
which is just one of the terms required to make $W_{\mu\nu}$
gauge invariant.

The contribution of Fig. 2c can be calculated
in the same way and the result is complex conjugate of eq.
(41) with $\mu, \nu$ indices
interchanged. Combining the $h_1(x)$ term in eq. (41), and
eq. (44) and its conjugate, we have the chiral-odd part of the
spin-dependent nucleon tensor,
$$ \eqalign{    W_{[\mu\nu]}^{a+b+c} = & {M\over 2\nu}
           \sum_a e_a^2 h^a_1(x_B) \int dz {\hat e^a_1(z) \over z}
           \Big[i \epsilon^{\mu\nu\alpha\beta}
              p_{\alpha}S_{\perp\beta}  \cr &
            + {1\over zx_B} {1\over p\cdot p_\pi}
            ip_\pi^{\mu}\epsilon^{\nu\alpha\beta\gamma}
             S_{\perp\alpha}p_{\beta}p_{\pi\gamma}
             - {1\over zx_B} {1\over p\cdot p_\pi}
            ip_\pi^{\nu}\epsilon^{\mu\alpha\beta\gamma}
             S_{\perp\alpha}p_{\beta}p_{\pi\gamma}\Big]  \cr
            = & -i\epsilon^{\mu\nu\alpha\beta}
              q_\alpha S_{\perp\beta} {M\over 2\nu}
             \sum_a e_a^2 {h^a_1(x_B)\over x_B}
                 \int dz {\hat e^a_1(z) \over z}  }
\eqno(45) $$
which is explicitly gauge invariant.

Now we consider the contributions from Figs. 2d and 2e.
The calculations here parallel these for Figs. 2b and 2c,
and the final result for the chiral even part of the
nucleon tensor, including the contribution from Fig. 2a,
is,
$$ \eqalign{    W_{[\mu\nu]}^{a+d+e} = & {M\over 2\nu}
           \sum_a e_a^2 g^a_T(x_B) \int dz {f_1^a(z) \over z}
           \Big[i \epsilon^{\mu\nu\alpha\beta}
              S_{\perp\alpha}p_{\pi\beta}  \cr &
            +  {x_Bz\over p\cdot p_\pi}
            ip^{\mu}\epsilon^{\nu\alpha\beta\gamma}
             S_{\perp\alpha}p_{\beta}p_{\pi\gamma}
             -  {x_Bz\over p\cdot p_\pi}
            ip^{\nu}\epsilon^{\mu\alpha\beta\gamma}
             S_{\perp\alpha}p_{\beta}p_{\pi\gamma}\Big] \cr
             = & -i\epsilon^{\mu\nu\alpha\beta}
              q_\alpha S_{\perp\beta} {M\over 2\nu}
             \sum_a e_a^2 g^a_T(x_B)
                 \int dz \hat f_1^a(z) }
\eqno(46) $$

Adding eqs. (45) and (46) to the longitudinal-polarization
contribution, which is considerably easy to calculate,
we have the complete spin-dependent nucleon tensor,
$$       W^{\mu\nu} = -i\epsilon^{\mu\nu\alpha\beta}
              {q_{\alpha}\over \nu}
          [(S\cdot n)p_\beta \hat G_1(x,z) +
        {S_{\perp \beta}} \hat G_T(x,z)] \eqno(47) $$
The two structure functions are defined as,
$$ \eqalign{     \hat  G_1(x,z) & = {1\over 2} \sum_a e_a^2 g_1^a(x)
\hat f_1^a(z)  \cr
       \hat G_T(x,z) & = {1\over 2} \sum_a e_a^2 \Big [
            g^a_T(x) \hat f^a_1(z) + {h^a_1(x) \over x}
       {\hat e^a(z)\over z} \Big]  } \eqno(48)  $$

To isolate the spin-dependent cross section we take
the difference of cross sections with
left-handed and right-hand leptons,
$$         {d^2\Delta \sigma \over dE'd\Omega}
          = {\alpha_{\rm em}^2 \over Q^4} {E'\over EM_N}
            \Delta \ell^{\mu\nu}W_{\mu\nu}    \eqno(49) $$
where $Q^2 = -q^2$, $k = (E, {\bf k})$ and
$k' = (E', {\bf k'})$ are the incident and outgoing momenta
of the lepton, and $\Delta \ell^{\mu\nu}$
is the spin-dependent part of the lepton tensor,
$\Delta \ell^{\mu\nu} = - 1/2
          {\rm Tr}[\gamma^{\mu}\thru k'\gamma^{\nu}\gamma_5\thru k]
              = - 2i\epsilon^{\mu\nu\alpha\beta} q_{\alpha}
            k_{\beta}$.
It is convenient to express the cross section
in terms of scaling variables in a frame
where the lepton beam defines the $z$-axis
and the $x-z$ plane contains the nucleon polarization vector,
which has a polar angle $\alpha$.
In this system, the scattered lepton has polar angles
$(\theta, \phi)$ and therefore the momentum transfer ${\bf q}$
has polar angles $(\theta, \pi-\phi)$. Defining a conventional
dimensionless variable $y=1-E'/E$, we can write the cross
section as
$$ \eqalign{      {d^4 \Delta \sigma \over dxdydzd\phi}
        = {4\alpha^2_{\rm em} \over Q^2}
         \Big[& \cos\alpha (1-{y\over 2}) \hat G_1(x,z) \cr
            + &\cos\phi\sin\alpha\sqrt{(\kappa -1)(1-y)}
         \left(\hat G_T(x,z) - \hat G_1(x,z)(1-{y\over 2})\right)\Big] }
  \eqno(50) $$
where $\kappa = 1 + 4x^2M^2/Q^2$ in the second term
signals the suppression by a factor of $1/Q$ associated with
the structure function $\hat G_T$. The existence of
of $\hat G_1$ in the same term is due to a small
longitudinal polarization of the nucleon when its spin
is perpendicular to the lepton beam.

Equation (50) is one of our main results. As a check, we multiply
by $z$, integrate over it and sum over all hadron species.
Using the well-known momentum sum rule,
$$  \sum_{\rm hadrons} \int dz z \hat f_1^a(z)  = 1 \eqno(51) $$
and the sum rule,
     $$   \sum_{\rm hadrons}  \int dz \hat e^a_1(z)  = 0  \eqno(52) $$
which is related to the fact that the chiral condensate
vanishes in the perturbative QCD vacuum,
we get,
$$ \eqalign{     {d^3 \Delta \sigma \over dxdyd\phi}
        = {4\alpha^2_{\rm em} \over Q^2}
         \Big[&\cos\alpha (1-{y\over 2}) g_1(x) \cr
            +& \cos\phi\sin\alpha\sqrt{(\kappa -1)(1-y)}
         \left(g_T(x) - g_1(x)(1-{y\over 2})\right)\Big] }  \eqno(53) $$
where  $ g_1(x) = {1\over 2}\sum_a e_a^2 g^a_1(x)$ and
$ g_T(x) = g_1(x) + g_2(x) = {1\over 2}\sum_a e_a^2 (g^a_1(x) + g^a_2(x))$
are the two conventional spin structure functions. The above
result coincides with the same quantity in ref. 12
if one neglects the terms of order $1/Q^2$ in the latter.
The parallelism between the inclusive and semi-inclusive cross sections
suggests that the both quantities can be extract from the same
set of experiment.

In using eq. (50) to analyze experimental data, a lower cut
on $z$ must be made to ensure the detected particles
emerging from the current fragmentation region.
To enhance statistics one can integrate $z$ over a
region. By varying $\phi$, we
can separate out the following combinations
of structure functions,
$$ \eqalign{      \int G_1 dz & = {1\over 2}\sum_a e_a^2g_1^a(x) N^a_\pi  \cr
              \int G_T dz & = {1\over 2}\sum_a e_a^2 [g_T^a(x) N^a_\pi
               + {h_1^a(x) \over x} E^a_\pi] } \eqno(54) $$
where $ N^a_\pi =\int dz f_1^a (z) $ is the
pion multiplicity of the quark jet with flavor $a$ and
$E_\pi^a = \int dz {e^a(z)/ z}$.

\goodbreak\medskip
\noindent
{\bf 2. Spin-1/2 Baryon Production}
\smallskip\nobreak

In this subsection we study deep-inelastic
scattering of unpolarized
lepton beam on transversely polarized nucleon target,
focusing on spin-1/2 baryon production from
quark fragmentation.
The spin effects in the scattering can be unravelled
through measuring the polarization of the
produced baryon. This can be done
for an unstable hyperon by measuring angular distribution
of its decay product. The process was first
studied in ref. 13. Here we include a formula
for the spin-dependent cross section in the
lab frame.

The process can be described as in Fig. 2(a),
except the produced pion is replaced here by
a spin-1/2 baryon. From eq. (14), we find
the spin-dependent piece of the fragmentation
density matrix,
$$          \hat M_B(z) = {\hat h_1(z)\over z}
       \gamma_5\thru S_{B\perp}\thru p_B
   + ... \eqno(55) $$
where $p_B = z\nu n$ and $S_B$
are the momentum and polarization of the baryon, respectively.
Thus the spin-dependent nucleon tensor
is,
$$ \eqalign{    W^{\mu\nu} = &
            - {1\over 2\nu} \sum_a e_a^2 h^a_1(x) {\hat h^a_1(z) \over z}
              \Big[ (S^\mu_\perp S^\nu_{B\perp}
              + S^\nu_\perp S^\mu_{B\perp}) p\cdot p_B
             \cr & + (p^\mu p_B^\nu+p^\mu p_B^\nu - g^{\mu\nu}p\cdot p_B)
                S_\perp\cdot S_{B\perp} \Big] }  \eqno(56) $$
Contracting it with the unpolarized lepton tensor,
$\ell_{\mu\nu} = {1\over 2}{\rm Tr}[\gamma_\mu\thru k\gamma_\nu\thru k']$,
we have,
$$       \ell^{\mu\nu}W_{\mu\nu} =
            -{4\over \nu} \sum_a e_a^2 h^a_1(x) \hat h^a_1(z)
              \Big[ S_\perp \cdot k S_{B\perp}\cdot k p\cdot p_B
             + k\cdot p k\cdot p_B
                S_\perp\cdot S_{B\perp} \Big]  \eqno(57) $$

Using the lab coordinate system defined in the last subsection
to simply (57), we find,
$$        \ell^{\mu\nu}W_{\mu\nu} = -4Q^2{1-y\over y}\cos(\phi+\phi')
             {1\over 2}\sum_a e_a^2 h_1^a(x) \hat h_1^a(z)  \eqno(58) $$
where $\phi'$ is the azimuthal angle between ${\bf k'}$ and ${\bf S}_B$.
This produces the following spin-dependent cross section,
$$         {d\Delta \sigma \over dxdydzd\phi}
          = - 4{\alpha_{\rm em}^2\over Q^2} {1-y \over y}
             \cos(\phi+\phi'){1\over 2}\sum_a e_a^2
             h_1^a(x) \hat h_1^a(z)  \eqno(59) $$
This expression reaches maximum if ${\bf S}$ and ${\bf S}_B$ are
the mirror imagine of each other with respect to
the scattering plane.

\goodbreak\medskip
\noindent
{\bf 3. Vector-Meson and Two-Pion Production}
\smallskip\nobreak

Here we consider the same set up for
deep-inelastic scattering as in the last subsection,
but focusing on vector meson, e.g. $\rho$,
and two-pion production production.
Our analysis shows that one can define
a single-spin asymmetry sensitive to the nucleon's
transversity distribution at the leading order
in $Q$, however, its magnitude depends also on the
unknown final-state interactions between the detected
particle(s) and spectators. Similar ideas have also been
proposed in refs. 14 and 15.

Let us first look at vector meson production. According
to our previous discussion on quark fragmentation
for a vector meson, there are two twist-two
{\it chiral-odd} fragmentation functions $
\hat h_1(z)$ and $\hat h_{\bar 1}(z)$.
The former describes the probability of
producing vector mesons in vector polarization
and the latter in tensor polarization.
If one can measure the vector polarization,
$\hat h_1(z)$ is an ideal choice for coupling
with the transversity distribution.
However, for the interesting case of $\rho$ meson
production, the only way to measure
polarization is through its
two-pion decay, which registers only
tensor polarization. Thus, it appears that
$\hat h_{\bar 1}(z)$ is the only choice
for coupling with the transversity
distribution. However, the size of this fragmentation
function depends on unknown final-state interactions.

If one is to measure asymmetry associated with
inclusive production of two pions, there are other
underlying processes which contribute besides the $\rho$ decay,
for instance, the interference production of two pions in their
relative $s$ and $p$ waves. The contribution depends
on the difference of the phase shifts. To include all
the contributions, we directly introduce quark fragmentation
functions for two-pion production,
$$  \hat  M(k,P_{2\pi},l) =   \int {d^4\xi \over (2\pi)^4}
           e^{-ik\cdot \xi}
       \langle 0|\psi(0)|\pi(l_1)\pi(l_2)X\rangle
       \langle \pi(l_1)\pi(l_2) X|\bar \psi(\xi)|0\rangle \eqno(60) $$
where $l_1$ and $l_2$ are momenta of two observed pion and
$P_{2\pi} = l_1+l_2$ and $l= (l_1-l_2)/2$ are the total and
relative momenta, respectively.

The contribution of two-pion fragmentation to
the nucleon tensor is,
$$      W_{\mu\nu} =
         {1\over 4\pi}\int {d^4P_{2\pi} \over (2\pi)^4}{d^4l \over (2\pi)^4}
           2\pi\delta(l\cdot P_{2\pi}/2) 2\pi\delta(4l^2 + P^2_{2\pi})
            dx {\rm Tr}
         \Big[M_N(xp,S_\perp)\gamma_{\mu}\hat M(xp+q,P_{2\pi},l)\gamma_{\nu}
          \Big]
          \eqno(61)  $$
where we have neglected the pion mass and
made collinear expansion for the intial quark momentum.
To proceed further, we make a restriction on the $l$
integrations such that $|l^2|<M^2$, where $M$ is a soft
scale on the order of $\Lambda_{QCD}$. Making a collinear
expansion for $P_{2\pi}$ and neglecting
higher-twist contributions, we have,
$$      W_{\mu\nu} = {1\over 4\pi} \int dx 2\pi \delta((xp+q)^2)
        \int {d^4l \over (2\pi)^4}  2\pi\delta(l\cdot p_{2\pi}/2)
            dx {\rm Tr} \Big[ M_N(xp, S_\perp)\gamma_{\mu}\hat
        M(p_{2\pi}/z,l)\gamma_{\nu}\Big]
          \eqno(62)  $$
where $p_{2\pi} =z \nu n$ and
the fragmentation density simplifies to,
$$   \hat  M(p_{2\pi}/z,l) =   \int {d\lambda \over 2\pi} e^{-i\lambda/z}
       \langle 0|\psi(0)|2\pi(p_{2\pi},l)X\rangle
       \langle 2\pi(p_{2\pi},l) X|\bar \psi(\lambda n_{2\pi})
       |0\rangle \eqno(63) $$
with $n_{2\pi} = p/(z\nu)$. For our purpose,
we make the following expansion for the density,
$$    \hat M(p_{2\pi}/z,l) = {\hat H_1(z,l) \over z}
       \gamma_5\thru S_{2\pi\perp} \thru p_{2\pi} +
          {{\hat F}_1(z,l)\over z}\thru p_{2\pi}+
       ... \eqno(64) $$
where $S^{\alpha}_{2\pi\perp} = \epsilon^{\alpha\beta\gamma\delta}
p_{2\pi\beta}n_{2\pi\gamma} l_{\delta}/|{\bf l}|$
and $S_{2\pi\perp}\cdot p_{2\pi} = 0$.
The fragmentation function $\hat H_1(z,l)$ is real according to
hermiticity and non-vanishing because of the final-state
interactions between $\pi$'s and $X$.

Substituting eq. (64) into eq. (62), we have,
$$ \eqalign{    W^{\mu\nu} = &
            - {1\over 2\nu} \sum_a e_a^2 h^a_1(x) \int {d^4l\over (2\pi)^4}
          2\pi \delta(l\cdot p_{2\pi}/2){\hat H^a_1(z,l) \over z}
              \Big[ (S^\mu_\perp S^\nu_{2\pi\perp}
              + S^\nu_\perp S^\mu_{2\pi\perp}) p\cdot p_{2\pi}
             \cr & + (p^\mu p_{2\pi}^\nu+p^\mu p_{2\pi}^\nu
             - g^{\mu\nu}p\cdot p_{2\pi})
                S_\perp\cdot S_{2\pi\perp} \Big]. }  \eqno(65) $$
For this, we can calculated the spin-dependent
part of the cross section,
$$         {d\Delta \sigma\over dxdydzd\phi}
            = {4\alpha^2_{\rm em} \over Q^2} {1-y\over y}
          {1\over 2} \sum_ae_a^2h_1^a(x)\int {d^4l\over (2\pi)^2}
  2\pi\delta(l\cdot p_{2\pi}/2)\hat H_1^a(z,l) \sin(\phi + \phi_\ell),
\eqno(66)
$$
where $\phi_{\ell}$ is the azimuthal angle between ${\bf k'}$
and ${\bf l}$.

\goodbreak\bigskip
\noindent
{\bf IV. CONCLUSION}
\medskip
\nobreak

In this paper, we define a number of low-twist
quark fragmentation functions by
analyzing the matrix elements of
quark bilinears in light-cone separations
and expanding them in terms of various Lorentz structures.
Some of these fragmentation functions are
chiral-odd and polarization-dependent, which
are not only interesting phenomenologically,
but also useful for describing non-perturbative
fragmentation processes.

In the examples of using the fragmentation functions,
we study measurement of the nucleon's transversity
distribution in deep-inelastic scattering, where chirality
conservation selects the class with odd chirality.
Fragmentation functions and parton distributions
are always coupled in cross sections, thus one can study both
in experiments by varying
$x$ and $z$ simultaneously. The facilities at CERN, HERA, and
SLAC are particularly useful for learning these
non-perturbative hadron observables.

\goodbreak
\bigskip
\line{\bf ACKNOWLEDGMENT \hfil}
\medskip
\nobreak
I thank Bob Jaffe for his collaboration on parton
fragmentation functions and J. Collins
for several useful conversations.

\goodbreak
\bigskip
\line{\bf REFERENCES \hfil}
\medskip
\item{1.}R. L. Jaffe and X. Ji, Phy. Rev. Lett.
{\bf 67} (1991) 552.
\medskip
\item{2.}R. L. Jaffe and X. Ji, Nucl. Phys. {\bf B375} (1992) 527.
\medskip
\item{3.}X. Ji, Phys. Lett. {\bf B284} (1992) 137.
\medskip
\item{4.}X. Ji, Phys. Lett. {\bf B289} (1992) 137.
\medskip
\item{5.}X. Ji, MIT CTP preprint No. 2141, 1992, to be
appeared in Nucl. Phys. B.
\medskip
\item{6.}J. P. Ralston and D. E. Soper, Nucl. Phys. {\bf B152} (1979) 109.
\medskip
\item{7.}J. C. Collins and D. Soper, Nucl. Phys. {\bf B194} (1982) 445.
\medskip
\item{8.}R. P. Feynman, {\it Photon-Hadron Interactions}, Benjamin,
Reading, MA. 1972.
\medskip
\item{9.}R. L. Jaffe and X. Ji, MIT CTP Preprint No. 2158, 1993.
\medskip
\item{10.}G. Alteralli and G. Parisi, Nucl. Phys. {\bf B126} (1977) 278.
\medskip
\item{11.}X. Ji and C. Chou, {\it Phys. Rev.} {\bf D42} (1990) 3637.
\medskip
\item{12.}R. L. Jaffe, {\it Comm. Nucl. Part. Phys.} {\bf 14} (1990) 239.
\medskip
\item{13.}X. Artru and M. Mekhfi, {\it Z. Phys.} {\bf  C45} (1990) 669.
\medskip
\item{14.}A. V. Efremov, L. Mankiewicz, and N. A. Tornqvist,
{\it Phys. Lett.} {\bf  B284} (1992) 394.
\medskip
\item{15.}J. Collins, {\it Nucl. Phys.} {\bf B396} (1993) 161.

\goodbreak
\bigskip
\line{\bf FIGURE CAPTIONS \hfil}
\medskip
[For hard copies of the figures for this paper, send email to
ereidell@marie.mit.edu.]
\medskip
\item{Fig.~1:} Processes in which a quark changes
its chirality: a) a generic soft QCD process,
b) mass insertion, c) the Drell-Yan scattering,
and d) the quark fragmentation.
\medskip
\item{Fig.~2:} The twist-two and twist-three cut diagrams
for single-pion production in deep-inelastic scattering.
\par
\vfill
\end